Filtration method characterizing the reversibility of colloidal fouling layers at a membrane surface: analysis through critical flux and osmotic pressure

Benjamin Espinasse, Patrice Bacchin[*], Pierre Aimar.

31062 Toulouse cedex 9

France.

Corresponding author:

Tel :+33 5 61 55 81 63

Fax : +33 5 61 55 61 39

bacchin@chimie.ups-tlse.fr

*Abstract:*

A filtration procedure was developed to measure the reversibility of fouling during cross-flow filtration based on the square wave of applied pressure. The principle of this method, the apparatus required, and the associated mathematical relationships are detailed. This method allows for differentiating the reversible accumulation of matter on, and the irreversible fouling of, a membrane surface. Distinguishing these two forms of attachment to a membrane surface provides a means by which the critical flux may be determined. To validate this method, experiments were performed with a latex suspension at different degrees of destabilization (obtained by the addition of salt to the suspension) and at different cross-flow velocities. The dependence of the critical flux on these conditions is discussed and analysed through the osmotic pressure of the colloidal dispersion.

Keywords: Critical flux, ultrafiltration, colloids, membrane, irreversibility, fouling, osmotic pressure.

## 1 Introduction

As the price of membranes continues to decrease, operating costs are representing a larger share of the total cost of membrane processes and are limiting the development of membrane technologies. In membrane filtration, these costs are primarily linked to fouling in two points: one is relative to the direct reduction of productivity and the other is linked to the selectivity of the operation. The formation of a deposit on the membrane surface generally changes its properties. This is an important problem for applications that are very sensitive to the surface properties, as in food and pharmaceutical industries[1]. When a natural dispersion is ultra-filtered, fouling is often the consequence of the concentration of colloids [2] (macromolecules or sub-micronic particle). Now, colloid dispersions exhibit a specific behaviour because of surface interactions. These surface interactions are multiple in nature (electrostatic repulsion, Van-der-Waals attraction or hydrophobic-philic interaction) and have different interaction lengths; thus leading to a complex system when colloidal dispersions are concentrated. It has been shown [3] that the accumulation of matter on the membrane surface can be characterised with regard to its reversibility. A reversible accumulation will "disappear" when the transmembrane pressure (TMP) is decreased, while an irreversible one (with the form of a deposit or a gel) will remain on the membrane surface when the pressure is released. In the case of irreversible fouling, the method for removing the deposit layer is mechanical or chemical cleaning. The limit between reversible and irreversible fouling often appears to operators when they increase the permeation flux; the threshold value for which the reversible accumulation turns into an irreversible one can be called a critical flux. It was defined in the literature in 1995 [4, 5] as the flux above which an irreversible deposit appears at the membrane surface. Its experimental determination is then of practical importance to optimize the operating conditions of a membrane process [3].

Mass accumulation on a membrane surface can be reversible if the matter is accumulated but stays dispersed, or does not stick to the surface (a reduction in the applied pressure reduces the concentration polarisation by diffusion). The osmotic model is well accepted to establish the J versus TMP relationship. However, the concept of osmotic pressure is well established for molecules and salts (and their limiting effects on nanofiltration and reverse osmosis) but not as much for colloidal dispersions in processes such as ultrafiltration. A number of studies have experimentally given evidence of the osmotic pressure of colloidal dispersions [6] with various measurement methods. Different authors have linked the osmotic pressure to the properties of the colloids[7, 8, 9] . The main difference between molecular and particulate osmotic pressure lies in the range of pressure (much lower for particles). Colloidal systems such as dispersions of latex particles can "resist" compression and "create" an osmotic pressure from around 0.1 bar to a few bars [10] according to the particle size and their inter-particle interactions. For some critical value of the osmotic pressure (or at the associated critical volume fraction, generally around 0.4 - 0.6 bar), the compressed dispersion is no longer stable and particles make an irreversible state transition from a dispersed to a condensed phase [11] where particles are all in contact with each other. This irreversible transition from a dispersed to a condensed phase can occur at the membrane surface when the permeate flux is high enough, resulting in a deposit or gel layer formation at the membrane surface. Consequently, deposit layer formation is intrinsically linked to the osmotic pressure and to the permeate flux that leads to aggregation of the particles on the membrane surface (i.e., the critical flux) [5].

The critical flux was defined in 1995 by two different teams both theoretically and experimentally [4, 5, 12]. Theoretically the critical flux was defined as the _flux above which an irreversible deposit appears at the membrane surface_. For colloidal suspensions this phenomenon is generally a balance of particle-particle or particle-membrane repulsive forces and permeate drag forces. Above a given value of flux, when the repulsive forces are overcome by the permeate drag forces, a deposit forms on the membrane surface and creates an additional resistance to the permeate flow through the membrane.

Based on a theoretical force balance approach, when the critical flux is overcome a decrease in pressure will not lead to a spontaneous decrease in resistance. Howell and al. [13] confirmed the existence of a critical and sub-critical flux where, at a constant flux, the pressure remains constant with time : no deposit appears for these operating conditions on the membrane surface, regardless of the duration of filtration. On the other hand, setting a flux above the critical flux will lead to an increase in pressure with time until a steady state condition has been reached. The concept of critical flux is also used in membrane bioreactor studies to determine the filtration regime by analyzing the variations in pressure for constant flux experiments [14]. Researchers often consider the critical flux to have been reached when the pressure cannot attain steady state with time for that given flux [15]. Other investigators[16], in contrast, measured the transmembrane pressure necessary to maintain a given permeation flux across a microfiltration membrane for silica suspensions by using a flux stepping method (Fig. 1).

These experiments determined two kinds of critical flux: the weak form (to separate non linear pressure-flux variations to linear variations with a slope being inferior to the one of the water flux) and the strong form (to separate non linear pressure-flux variations to linear variations similar that the one reached for water flux).

Other techniques are also used to detect deposit formation: evolution of concentration in the suspension [17] in a batch operation can allow estimation of the accumulated material via mass balance. Direct microscope observation [18] has also been used to study the evolution of a deposit on a membrane surface for particles having a size larger than a few microns, but as an indirect method, the accuracy and relevance are unsatisfactory.

The aim of this work is to present a square wave filtration method to determine the extend of reversibility of fouling during membrane filtration (the general principle has been briefly presented in [19]). The primary goal of this method is to give a more accurate value of the critical flux than may be measured using other techniques. The improvement is based on the use of a square wave filtration method which makes possible the determination of the accumulation irreversibility all along the permeate flux range. This method determines the reversible and irreversible part of flux reduction due to accumulation of matter. The reversible part is described by an osmotic pressure contribution whereas the irreversible part is analysed as an hydraulic resistance (deposit or gel-like layer). This method allows defining, on a pragmatic way, the critical flux and is here validated for different conditions of filtration and colloïdal stability.

## 2  Materials and Methods

### 2.1  Filtration rig

The ultrafiltration set-up used in this study is shown in Fig. 2. The pressure is controlled by a Current-to-pressure Transducer (CPT, Rosemount, Baar, Switzerland) with an accuracy of 0.02 Bar and regulated with a PID (proportional-integral-derivative) regulator. The filtration rig is temperature controlled at 25°C and the permeation flux is measured with an electronic balance (Adventurer, Ohaus, Nanikon, Switzerland) linked to a computer. Crossflow is measured with a flow-meter (Promag A, Endress-Hausser, Reinach, Switzerland) with a precision of 3%.

The ultrafiltration module contains one inner skinned Carbosep tubular membrane (Orelis, Miribel, France). Its molecular weight cut off was 15 kDa.

The total active membrane surface was 0.0226 $m^2$, the length is 1m and the hydraulic diameter is 6 mm. In this rig, the flow velocity can be varied from 0.30 $m.s^{-1}$ to 1.27 $ms^{-1}$. The Reynolds numbers associated with these flow rates are 1952 and 8458, respectively.

The concentration of the dispersion in the filtration rig remains constant at 0.7 g/L. To do that, the retentate is recycled and the volume of the rig is automatically refilled by a volume equivalent to the permeate volume. The refilling solution has the same ionic strength. The criteria to end a pressure step is a variation of permeate flux inferior to 2% per hour. In this work, the time necessary to reach the equilibrium is ranged from 30 minutes to 4 hours. The pressure in the system is obtained with compressed air that pressurizes the feed tank. The pressure is accurately regulated in the rig through a current to pressure transducer controlled with a computer software interface.

*2.2 Latex suspension*

Latex particles used in these experiments are hard PVC spheres with a zêta potential ranging from -60± 2 mV at pH 3.5 to -85 ± 2 mV at pH 9. In the working conditions (pH 6-7) and when no salt is added, the average zeta potentiel is -71 ± 2 mV and the mean diameter is of 115 nm (Zetasizer 4, Malvern Instruments, Worcestershire, UK).

| Ionic strength (M) | Zeta potential (mV) |     |
|---|---|---|
| 0 |     | -71± 2 |
| 0    | -49 ± 1 | -71 ± 2 |
| $10^{-4}$ | -57 ± 1 | -76 ± 2 |
| $10^{-3}$ | -61 ± 1 | -71 ± 2 |
| $10^{-2}$ | -76 ± 1 | -80 ± 2 |
| $10^{-4}$ | -76 ± 2 |     |
| $10^{-3}$ | -71 ± 2 |     |
| $10^{-2}$ | -80 ± 2 |     |

**Table 1: Evolution of zeta potential with the ionic strength**

The particles are 100% rejected by the membrane. The salt concentration, in all experiments was below the critical coagulation concentration (0.1M in KCl) to avoid particle aggregation in initial dispersion. In this work, salt acts on the stability of particles that decreases when salt concentration increases. To have an experimental evidence of this effect, the osmotic pressure of the latex suspension was measured by a chemical compression method. Details of this method can be find in literature [8, 10, 20]. The results for an ionic strength of $10^{-3}$ M of KCl are presented in Fig. 3. where a classical increase in osmotic pressure with the latex volume fraction is observed. In this work, the increase in osmotic pressure relates the increase of the colloidal repulsive electrostatic interactions as particles are brought closer together. To illustrate this effect, these variations were proven to be sensitive to the inter-particle interactions by changing the ionic strength [20]: an increase in ionic strength leads to a decrease in osmostic pressure because of the reduction in repulsive interactions of the dispersion.

At a given volume fraction the particles are no longer stable and they turn from a dispersed to a condensed phase (the dispersion

| | | |
|---|---|---|
| 0 | -49 ± 1 | -71 ± 2 |
| $10^{-4}$ | -57 ± 1 | -76 ± 2 |
| $10^{-3}$ | -61 ± 1 | -71 ± 2 |
| $10^{-2}$ | -76 ± 1 | -80 ± 2 |

appears as a solid). This transition has been verified by re-dispersing 1 g of the compressed suspension into 200 ml of distilled water under agitation for 24 hr. Afterwards, the bulk suspension was analysed (by turbidity measurement) to evaluate the proportion of particles that re-dispersed. When no particles were re-suspended (black symbol in figure 3), it is concluded that the phase transition had occurred during the compression (the latexes are condensed during the osmotic pressure measurement test). This transition is linked to the concept of critical flux, which corresponds to an irreversible transition between a reversible to an irreversible deposit on the membrane.

## 3 Results and discussion

### 3.1 Principles and first results of the Square wave barovelocimetry method (SWB)

The principle of this filtration technique is to alternate stepwise the applied force (the pressure) with positive and negative variations, as presented in Fig. 4, and to continuously measure the permeate velocity. The U steps correspond to the upper steps and the L steps to the lower steps.

At each time step, a steady state permeate flux is reached. The stabilisation of the system (i.e., steady state) takes from a few minutes to several hours depending on system evolution. The method is similar to the electrochemistry method of "square wave voltammetry" and is defined as the square wave baroflumetry.

The interest of this technique is to evaluate the loss in flux between two steps of pressure: the flux is compared between step $L_n$ and $U_{n-1}$ which have the same pressure. The fouling associated to a decrease phenomena that took place at step $U_n$ is considered as totally reversible if the flux is the same at step $U_{n-1}$ and step $L_n$, and partly irreversible if not.

The classical J versus TMP filtration curve corresponds in Fig. 5 to the dotted line (upper pressure steps). The SWB method provides additional information with lower pressure steps that will allow for the deduction of the reversibility of the accumulated layer.

The hydraulic resistance of fouling layers can be classically determined through the integrated form of Darcy law:

$$J = \frac{\Delta P}{\mu (R_m + R_f)} \qquad [1]$$

The results are presented in terms of fouling resistance, Rf, over membrane resistance, Rm, in Fig. 6 as a function of the permeation flux on the y-axis in Fig. 5.

The Fig. 6 is read from the lower left point (start) and follows the arrows. At the beginning of filtration the fouling resistance increases with increasing pressure, but a decrease of pressure allows it to go back to the preceding resistance value; there is no hysteresis as the increase in resistance is reversible. Otherwise, an increase of resistance can be observed between steps n and n+1, such that resistance remains constant when the pressure is decreased at $L_{n+1}$ (the increase of resistance is consequently irreversible). The significance of this pattern is the appearance of an irreversible deposit on the membrane surface.

*3.2   Analysis of the filtration in term of irreversible and reversible resistance*

*3.2.1   Calculation of the irreversible resistance $R_{if}$*

From the representation $R_f/R_m$, it is possible to determine a grade of reversibility of the matter accumulated at the membrane surface: the term "reversible resistance" being here used to describe the portion of the fouling resistance that is eliminated with a decrease in pressure. The SWB method allows analysing the fouling reversibility by comparing the fouling resistance at a same TMP before and after an upper pressure step.

The irreversible resistance that appears for a upper pressure step $U_n$ can be reached by comparing the fouling resistance at step $L_n$ and $U_{n-1}$ (as defined in Fig. 4, 6 and 7) as follow:

$$\frac{r_{if,n}}{R_m} = \left.\frac{R_f}{R_m}\right|_{L_n} - \left.\frac{R_f}{R_m}\right|_{U_{n-1}} \qquad [2]$$

Where $r_{if,n}$ is the irreversible fouling resistance relative to the step n. When $R_f$ at step $L_n$ equals $R_f$ at step $U_{n-1}$ the resistance observed at $U_n$ step is totally reversible. To calculate the value of the total $R_{if}$, at a given pressure step, all step resistances $r_i$ are summed. The fouling resistance at n is the sum of the resistance measured at the previous steps:

$$R_{if} = \sum_n r_{if,n} \qquad [3]$$

Where $R_{if}$ is the total irreversible resistance. If the fouling is totally irreversible then $R_f\big|_{L_n} = R_f\big|_{U_{n-1}}$. An analysis of the $R_f$ and $R_{if}$ allows then the differentiation of the reversible and irreversible part of fouling in regard to a decrease of pressure all along the permeation flux range.

*3.2.2 Calculation of reversible resistance $R_{rf}$*

From the data previously calculated, in comparison with water flux, it is possible to deduce the contribution of osmotic pressure in the total fouling resistance. At each step, one can find the resistance associated with the reversible resistance, named $R_{rf}$ as follows:

$$\frac{R_{rf}}{R_m} = \frac{R_f}{R_m} - \frac{R_{if}}{R_m} \qquad [4]$$

The reversible and irreversible resistances deduced from the SWB experiment (in Fig. 6) are plotted on Fig. 8.

At low flux, the total resistance is low (less than 10% of the membrane resistance) and totally reversible. For a given flux (by definition: the critical flux) appears an irreversible resistance which is rapidly growing above this critical value.

*3.3 Analysis in terms of irreversible fouling resistance and osmotic pressure*

The reversible accumulation of matter (that goes back into suspension after a decrease of pressure) is related to the polarisation layer and induced osmotic pressure, which act as a force opposite to the applied pressure. In this case, the reversible resistance associated with the polarization concentration layer can be treated like a term of osmotic pressure:

$$J = \frac{\Delta P - \Delta \Pi}{\mu(R_m + R_{if})} = \frac{\Delta P}{\mu(R_m + R_{if} + R_{rf})} \qquad [5]$$

The osmotic pressure at the membrane, $\Pi_m$, can be deduced as follows assuming that the membrane is totally retentive:

$$\Delta\Pi = \Pi_m = \Delta P - J\mu R_m \left(1 + \frac{R_{if}}{R_m}\right) \quad [6]$$

Eq. [6] allows one to calculate the value of the osmotic pressure along the length of the filtration process with the square wave technique, as shown in Fig. 9. The slight increase in the reversible resistance (as presented in Fig. 8) leads to a more pronounced increase in the osmotic pressure at the membrane with flux (Fig. 9).

The osmotic pressure at the membrane increases due to the evolution of the polarisation layer thickness on the membrane surface (this accumulation is fully reversible if the pressure is decreased). Beyond a given flux (~1.4 $10^{-5}$ m.s$^{-1}$) is reached, the polarisation layer is thick enough for particles next to the membrane surface aggregate and lead to the creation of an irreversible layer (deposit-like). Beyond this point, the irreversible resistance increase denotes that the deposit layer thickness and/or the surface of the membrane covered by the deposit are increasing. We also observe that the osmotic pressure is increasing above the flux where the first reversible fouling is noticed. That may be due the growth of the polarization layer in zone along the membrane where there is no deposit ; the deposit layer is probably not covering all the membrane length when the critical flux is reached as it will be discussed in next section. The critical flux, the evolution of the irreversible resistance and the osmotic pressure are function of the particle stability and crossflow velocity.

*3.4    Effect of operating conditions*

In this part, the sensitivity of the SWB technique is presented for experiments at different cross-flow velocities. The degree of particle destabilization is discussed through the evolution of the irreversible resistance, the osmotic pressure and the critical flux.

*3.4.1    Effect of crossflow velocity*

Results from SWB experiments that are presented in Fig. 10 were performed at crossflow velocities of 0.3m.s$^{-1}$ and 0.79 m.s$^{-1}$. These velocities correspond to a laminar flow zone. With these membranes, the transition between laminar and turbulent was measured to be above 0.79 m.s$^{-1}$ [21] The latex concentration was 0.6 g.L$^{-1}$ and the salt concentration 10$^{-3}$ M KCl.

The first notable point (Fig. 10) is that the irreversible resistance starts to increase at higher permeation flux at high crossflow velocity. The second point is that the slope of irreversible resistance $R_{if}/R_m$ decreases with an increase of the crossflow velocity. Those observations are directly related to the evolution of the polarisation layer and have been described in the literature:  critical and limiting flux increases with increasing crossflow velocity. Osmotic pressure is almost identical for a small flux but the increases is less for higher crossflow velocities.

When the irreversible resistance appears, the osmotic pressure at the membrane is around 7000-12000 Pa at 0.3 m.s$^{-1}$ and 13000-15000 Pa at 0.79 m.s$^{-1}$. These values are smaller than those determined for the critical transition (16000-23000 Pa) measured with dialysis bags (Fig. 3). This difference may be explained by the fact that the value of the osmotic pressure at the membrane, as determined with the SWB method, corresponds to an averaged value of osmotic pressure along the membrane channel. If locally the osmotic pressure corresponding to the condensation of colloids appears at the membrane outlet (i.e. where the mass boundary layer is thicker), the average value of the osmotic pressure along the membrane length when critical conditions are reached will be less than the critical osmotic pressure. The SWB results show that the averaged osmotic pressure at the membrane increases with the crossflow velocity. This behaviour can be explained by a different shape of the polarisation layer along the channel for different cross flow velocities. Increasing the crossflow velocity may lead to the formation of a more homogenous polarisation layer. This could explain that at high crossflow velocities, the average osmotic pressure is close to the critical osmotic pressure measured with dialysis bags. The interpretation of the interdependence of the osmotic pressure and the deposit will need more efforts in the future with different systems (hollow fibres, flat membrane with different length, etc…) to be able to draw more generic conclusion on the dependence of osmotic pressure within geometry of modules and polarisation layers.

*3.4.2 Effect of the stability of latex dispersion*

The evolution of the irreversible resistances and the osmotic pressure as a function of the permeation flux is presented on Fig. 11 at a crossflow velocity of 0.6 m.s$^{-1}$ and a latex concentration of 0.6 g.L$^{-1}$. The latex dispersion is more or less destabilized by salt addition: the addition of salt reduces particles interactions and consequently leads to the apparition of a critical flux at lower flux when particles are less stable (i.e. at 10$^{-3}$ M in KCl). The increase in critical flux is accompanied by a decrease in the fouling rate (slopes of $R_{if}/R_m$ vs. flux are less important when salt concentration is reduced). These results confirm the importance of the inter-particles interactions in the fouling mechanisms.

The osmotic pressure at the membrane is less important when filtering more stable particles (no salt added). The hypothesis that can be formulated from this is that the volume fraction of particles at the membrane surface is less important with stable particles because they have a better resistance to compression by a more important collective diffusivity. These experimental results confirm theoretical calculations of cross flow filtration that have been discussed in previous article [22]. The variations of critical flux values with cross flow velocity and salt concentration are summarized in Fig. 12.

As expected, the values of critical flux decrease with the ionic strength, and increases with the increase of the crossflow velocities. The variation of the critical flux at two different crossflow velocities 0.3 and 1.27 m.s$^{-1}$ is really important in regard to the observed values in industrial ultrafiltration processes.

## 4  Conclusions

Using the square wave baroflumetry (SWB) technique to analyze membrane fouling data provides a quantitative assessment of the evolution of reversible resistance (osmotic pressure at the membrane) and irreversible resistance with the permeation flux (and the trans-membrane pressure). It allows an accurate determination of the critical flux, with regards to the irreversibility of fouling layers. Furthermore, the analysis of fouling in regard to the operating conditions allows quantifying the effect of crossflow velocity and particle stability on the growing of the polarized layer and the deposit formation. The measured variation with the operating conditions of the critical flux and the osmotic pressure confirm the importance of the colloidal interactions (or more precisely colloidal stability) on membrane fouling and its reversibility.

List of figures:

Fig. 1: Filtration procedure with flux stepping used in literature [4] to find the critical flux.

Fig. 2: Flow sheet of the filtration rig used for the critical flux determination.

Fig. 3: Results of measurements of the osmotic pressure vs. volume fraction of particles of PVC latex in deionised water and [KCl] = $10^{-3}$ M.

Fig. 4: Principle of the square wave technique; pressure and flux vs. time; upper and lower steps.

Fig. 5: Flux vs. TMP for the "square wave" technique; PVC latex, Qc = 0.79 m.s$^{-1}$, [KCl] = $10^{-3}$ M.

Fig. 6: Evolution of total fouling resistance vs. pressure obtained with the "square wave technique".

Fig. 7: Detail of Fig. 6 showing a full pressure step in term of R/Rm vs. Flux.

Fig. 8: Evolution of reversible, $R_{rf}$, and irreversible, $R_{if}$, fouling resistance vs. flux.

Fig. 9: Evolution of the osmotic pressure at the membrane and the irreversible resistance vs. permeation flux.

Fig. 10: Evolution of resistances and decomposition in term of osmotic pressure and irreversible resistance for two crossflow velocities; [latex] = 0.6 g.L$^{-1}$ [KCl] = $10^{-3}$ M.

Fig. 11: Evolution of Rf/Rm and Rif/Rm and osmotic pressure for two different states of destabilization of particles; different ionic strength, crossflow velocity = 0.6 m.s$^{-1}$, [latex] = 0.6 g.L$^{-1}$.

Fig. 12: Evolution of critical flux vs. the ionic strength for two crossflow velocities.

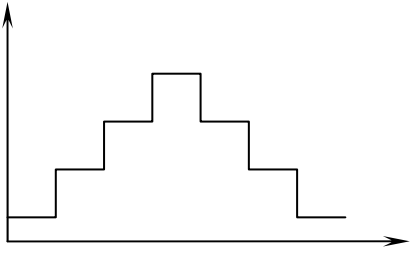

**Fig. 1: Filtration procedure with flux stepping used in literature [1] to find the critical flux.**

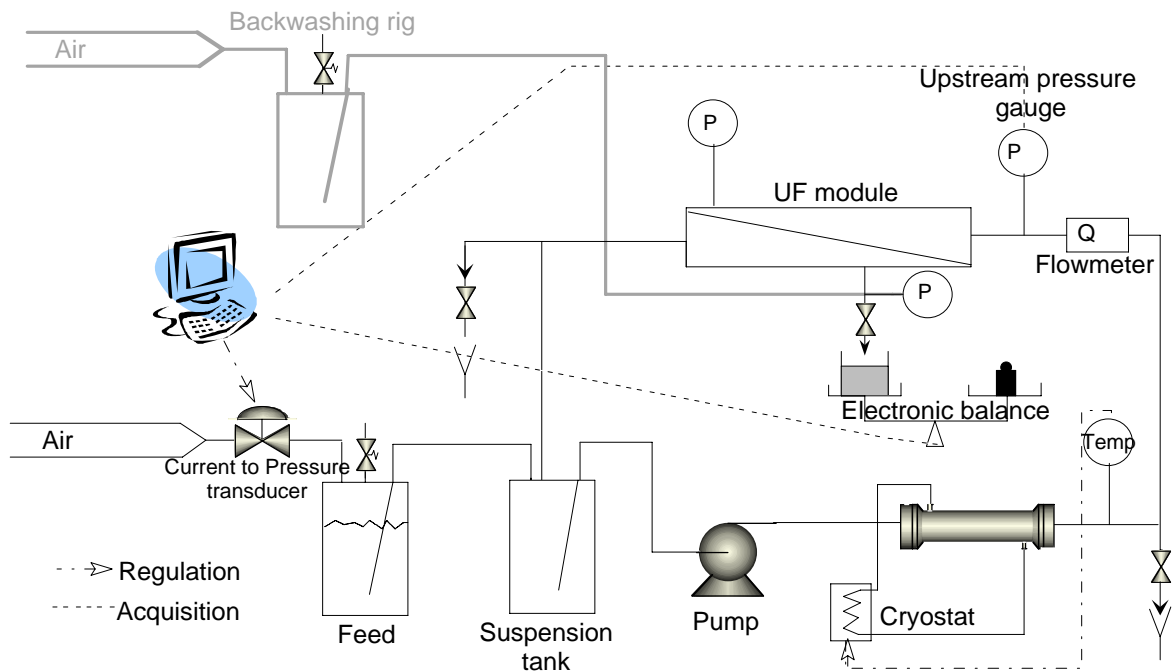

**Fig. 2: Flow sheet of the filtration rig used for the critical flux determination.**

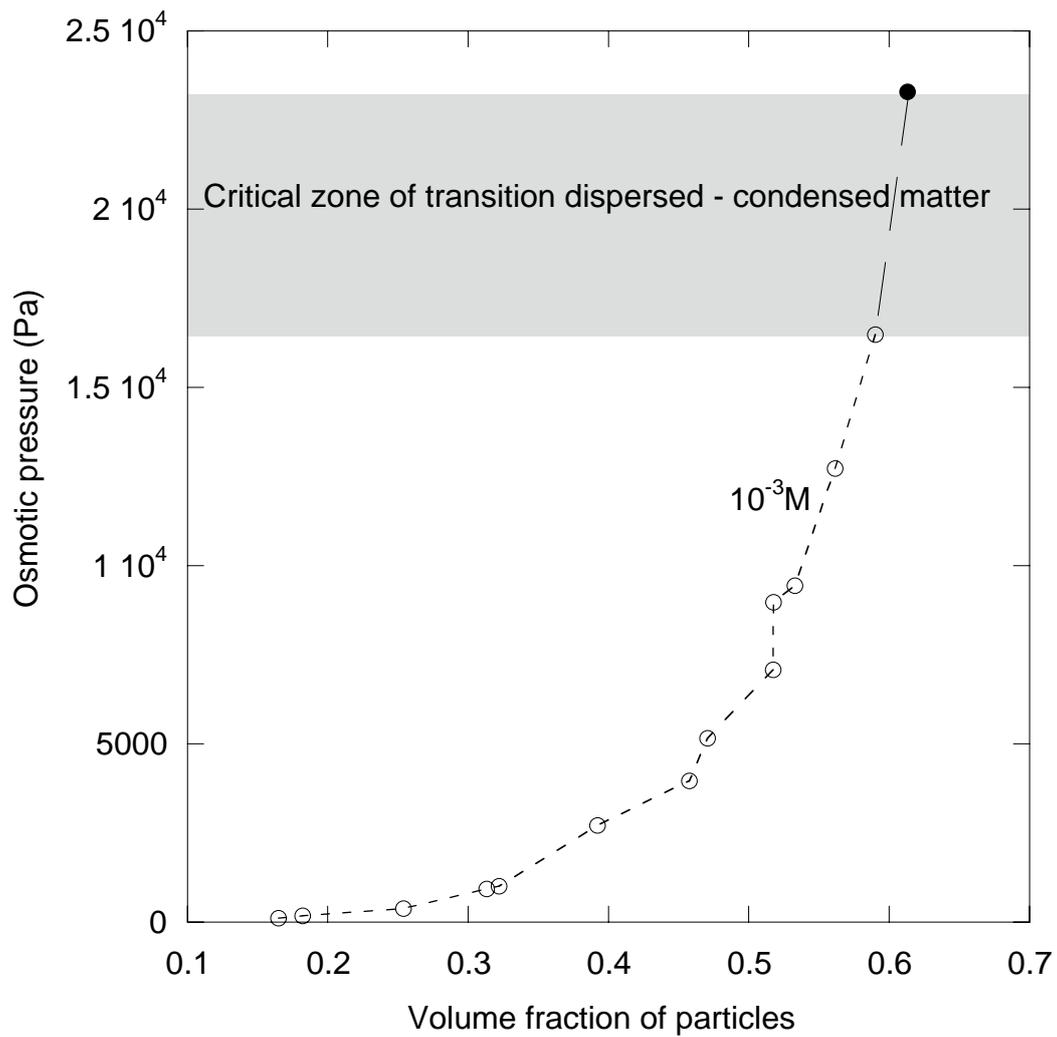

**Fig. 3: Results of measurements of the osmotic pressure vs. volume fraction of particles of PVC latex in deionised water and [KCl] = $10^{-3}$ M.**

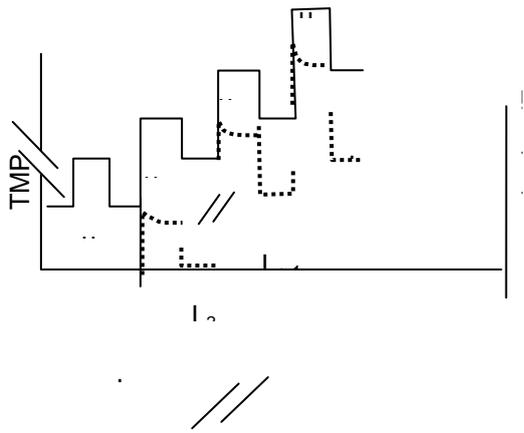

**Fig. 4: Principle of the square wave technique; pressure and flux vs. time; upper and lower steps.**

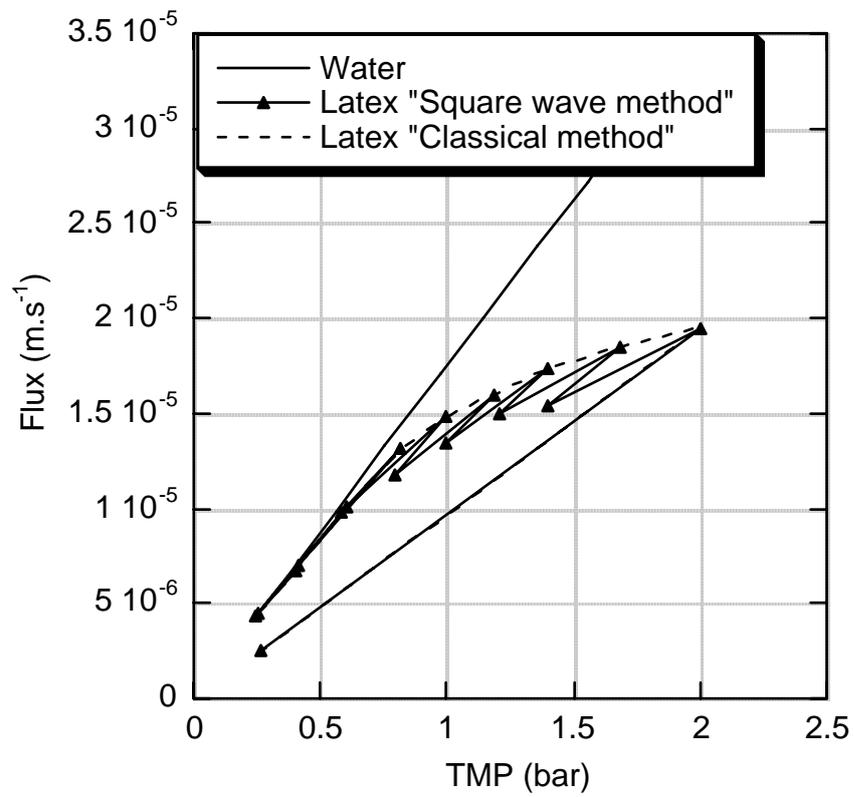

Fig. 5: Flux vs. TMP for the "square wave" technique; PVC latex, $Q_c = 0.79$ m.s$^{-1}$, [KCl] = $10^{-3}$ M.

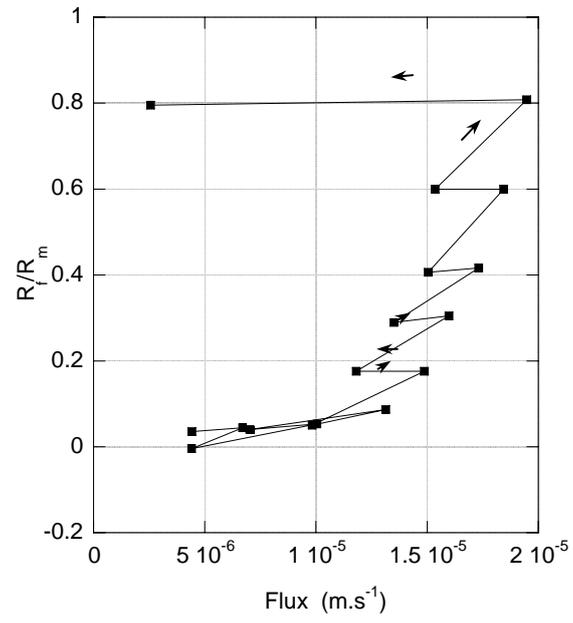

**Fig. 6: Evolution of total fouling resistance vs. pressure obtained with the "square wave technique".**

R/Rm

Zoom from Fig. 6

$I_n$  $U_n$

$I_{n-1}$  $U_{n-1}$

Flux (m.s$^{-1}$)

**Fig. 7: Detail of Fig. 6 showing a full pressure step in term of R/Rm vs. Flux.**

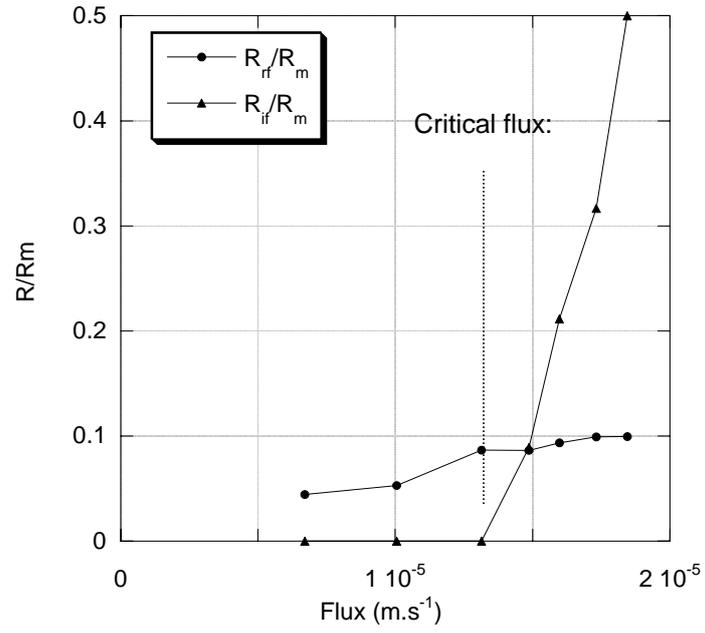

**Fig. 8: Evolution of reversible, R$_{rf}$, and irreversible, R$_{if}$, fouling resistance vs. flux.**

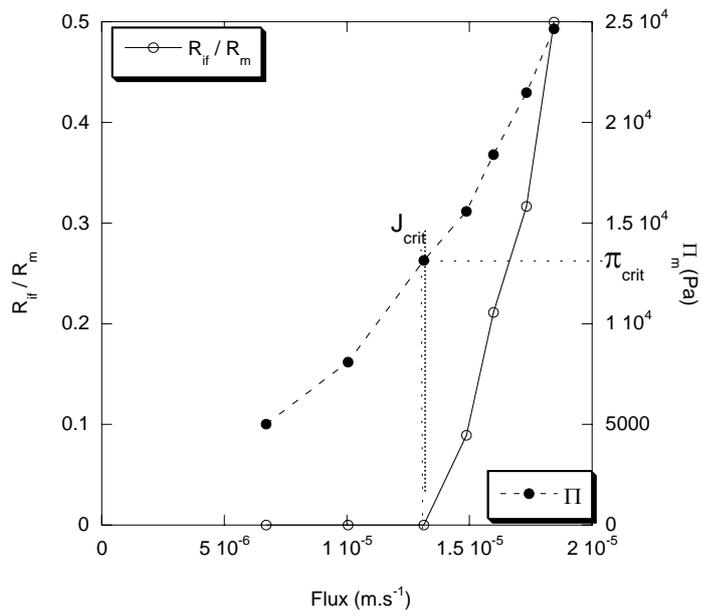

**Fig. 9: Evolution of the osmotic pressure at the membrane and the irreversible resistance vs. permeation flux.**

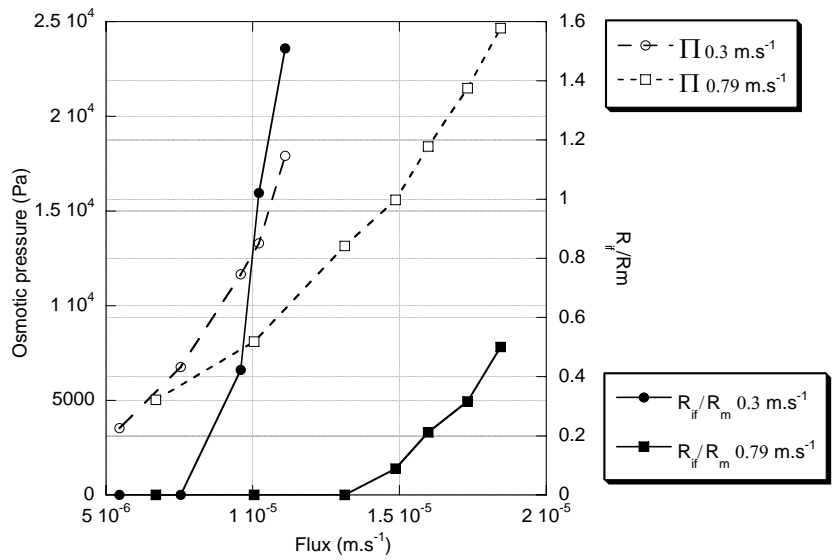

**Fig. 10: Evolution of resistances and decomposition in term of osmotic pressure and irreversible resistance for two crossflow velocities; [latex] = 0.6 g.L$^{-1}$  [KCl] = 10$^{-3}$ M.**

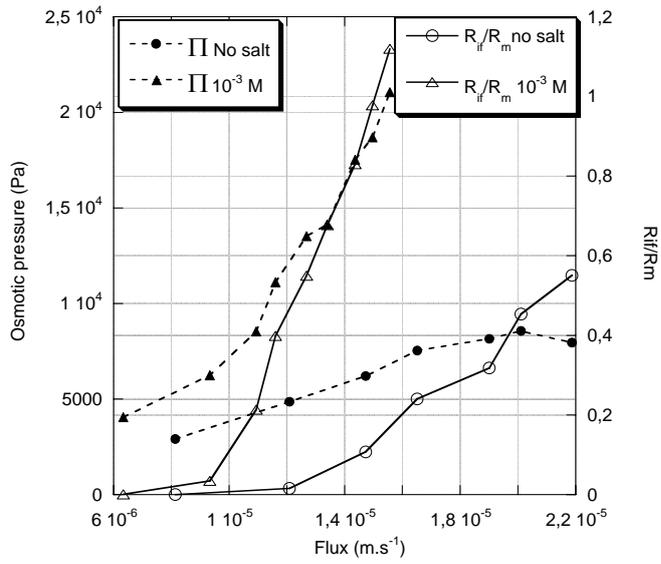

**Fig. 11: Evolution of Rf/Rm and Rif/Rm and osmotic pressure for two different states of destabilization of particles; different ionic strength, crossflow velocity = 0.6 m.s$^{-1}$, [latex] = 0.6 g.L$^{-1}$.**

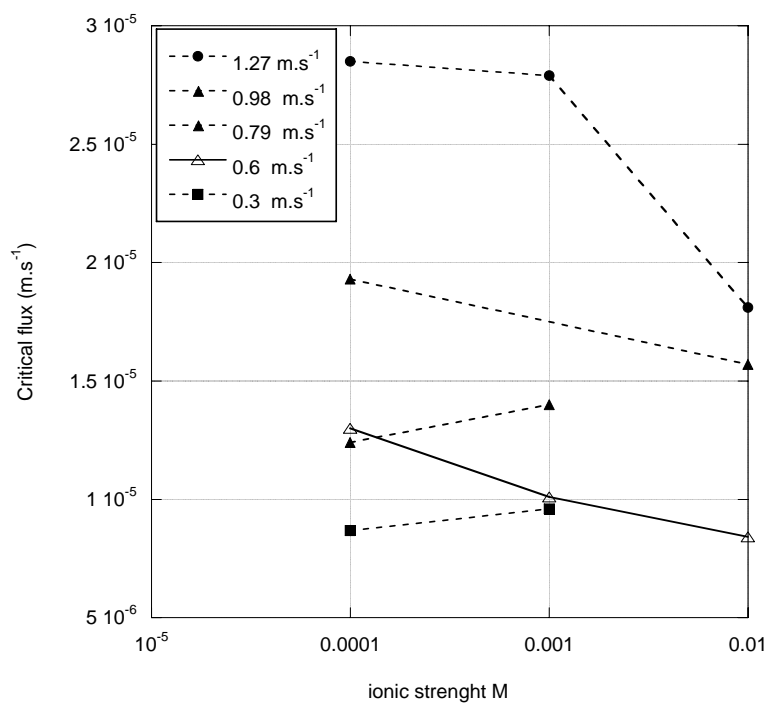

**Fig. 12: Evolution of critical flux vs. the ionic strength for two crossflow velocities.**